\documentclass[prl,amsmath,aps,twocolumn,superscriptaddress]{revtex4}
\usepackage{graphicx,xspace}
\usepackage{epsf}
\usepackage{float}
\usepackage{amsmath}
\usepackage{amssymb}
\usepackage[normalem]{ulem}
\input{epsf}
\begin{document}
\newcommand{\scalar}[2]{\left \langle#1\ #2\right \rangle}
\newcommand{\me}{\mathrm{e}}
\newcommand{\mi}{\mathrm{i}}
\newcommand{\dif}{\mathrm{d}}
\newcommand{\period}{\text{per}}
\newcommand{\free}{\text{fr}}
\newcommand{\mq}[2]{\uwave{#1}\marginpar{#2}} 

\title{Numerical study on Schramm-Loewner Evolution in nonminimal conformal field theories.}
\author{Marco Picco}
\email{picco@lpthe.jussieu.fr}
\affiliation{LPTHE, Universit\'e Pierre et Marie Curie-Paris6 and 
               Universit\'e Denis Diderot-Paris7\\
               4 Place Jussieu, 75005 Paris, France}
\author{Raoul Santachiara} \email{santachi@lpt.ens.fr}
\affiliation{CNRS-Laboratoire de Physique Th\'eorique de l'Ecole Normale
  Sup\'erieure, 24 rue Lhomond, 75231 Paris, France.} 
  
\begin{abstract}
  The Schramm-Loewner evolution (SLE) is a powerful tool to describe
  fractal interfaces in 2D critical statistical systems, yet the
  application of SLE is well established for statistical systems
  described by quantum field theories satisfying only conformal
  invariance, the so-called minimal conformal field theories
  (CFTs). We consider interfaces in $Z(N)$ spin models at their
  self-dual critical point for $N=4$ and $N=5$.  These lattice models
  are described in the continuum limit by nonminimal CFTs where the
  role of a $Z_N$ symmetry, in addition to the conformal one, should
  be taken into account. We provide numerical results on the fractal
  dimension of the interfaces which are SLE candidates for nonminimal
  CFTs. Our results are in excellent agreement with some recent
  theoretical predictions.
\end{abstract}

\maketitle 
{\it Introduction}--- The description of phase transitions
in terms of geometrical objects is a long-standing problem
\cite{Duplantier} which has provided a different conceptual framework
to study critical phenomena. In this respect, the two dimensional (2D)
systems are particularly interesting as an extensive variety of
theoretical tools is available. In particular, the approach based on
the so called Schramm-Loewner evolutions (SLEs), which are growth
processes defined via stochastic evolution of conformal maps, has been
proven an efficient tool to study fractal shapes in 2D critical
statistical systems and unveiled geometrical properties of critical
systems that were missing before
\cite{Walter,Cardy_review,Bernard_review}.

The SLE approach has been applied to different problems as the
critical percolation \cite{Smirnov}, the domain boundaries in magnetic
systems at the phase transition \cite{Bernard_review} or the 2D
turbulence \cite{Turb}. The theoretical ideas behind this approach
often combines the probability theory, the complex analysis and the
quantum field theory.  The conformal field theories (CFTs) play a key
role for understanding the universal properties of 2D systems
\cite{DiF}. If SLEs consider directly the geometrical characterization
of non-local objects, the CFTs focus on the computation of the
correlation function of local variables by fully exploiting the
symmetries of the system under consideration. The first solutions of
CFTs, the so called minimal CFTs, were constructed by demanding the
correlation functions to satisfy the conformal symmetry alone
\cite{BPZ}. So far the SLE interfaces have been identified and studied
in statistical models (critical percolation, self-avoiding walks, loop
erased random walks, etc.), which are described in the continuum limit by
minimal CFTs.  One of the most important results is the relation
between the SLEs and the minimal CFTs which has been worked out in
\cite{Bernard_connection1,Bernard_connection2,Bernard_connection3}. Yet,
there are other solutions of quantum fields theories which satisfy, in
addition to the conformal symmetry, additional symmetries. These
theories, called non-minimal CFTs, describe many condensed matter and
statistical problems characterized in general by some internal
symmetry such as, e.g., the $SU(2)$ spin-rotational symmetry in spin
chains \cite{Affleck} or replica permutational symmetry in disordered
systems \cite{Ludwig,DPP}. The connection between SLEs and non-minimal
CFTs has been first addressed in \cite{Rasmussen1,Rasmussen2}, where
the relation between stochastic evolutions and superconformal field
theory was investigated. More recently, the connection between SLE and
Wess-Zumino-Witten models, i.e. CFTs with additional Lie-group
symmetries, has been studied by very different approaches
\cite{Rasmussen3, Ludwig2}.  These results concern mainly some
particular properties of the CFTs under consideration which generalize
the ones on which the link between SLE and minimal CFT is
based. However, an interpretation in terms of the continuum limit of
lattice interfaces, necessary to give the SLE a physical meaning, was
missing.  In this respect, an interesting model is the $Z(N)$ spin
model (defined below) \cite{Raoul}, i.e. a lattice of spins which can
take $N$-values. The nearest-neighbor interaction defining the model
is invariant under a cyclic permutation of the states.  For $N=2$ and
$N=3$ one finds respectively the Ising and the three-state Potts
model. The phase diagrams of these $Z(N)$ spin models present
self-dual critical points \cite{Zama_lat2, Alcaraz,CardyZn} described
in the continuum limit by CFTs with $Z_N$ additional symmetries, the
so called $Z(N)$ parafermionic theories \cite{Zamo1}.  For $N\geq 4$
the parafermionic theories are non-minimal CFTs where the role of the
$Z_N$ symmetry beside the one of conformal symmetry must be taken
into account (for $N=2,3$ these theories coincide with minimal
models). In \cite{Raoul} the interfaces expected to be described in
the continuum limit by SLE have been identified on the lattice.
Further, combining CFT results with the idea, suggested in
\cite{Ludwig2}, of an additional stochastic motion in the internal
symmetry group space, the geometric properties of these interfaces was
predicted to be described by some specific SLE process.  In this
letter, we will investigate this model further and we will check the
prediction against numerical simulations for the self dual critical
$Z(4)$ and $Z(5)$ spin models.  We present the first numerical results
on critical interfaces on the lattice which are SLE candidates for non
minimal conformal field theories. Before presenting the model that we
simulate and the results we give some more definitions on SLEs.

{\it Schramm-Loewner evolution.}  Here we consider chordal SLE
which describes random curves joining two boundary points of a
connected planar domain.  For a detailed introduction to SLE, see
e.g. \cite{Walter,Cardy_review,Bernard_review}. The definition of SLE
is most conveniently given in the upper half complex plane
$\mathbb{H}$: it describes a fluctuating self-avoiding curve
$\gamma_{t}$ which emanates from the origin ($z=0$) and progresses in
a properly chosen time t. If $\gamma_{t}$ is a simple curve, this
evolution is defined via the conformal map $g_t(z)$ from the domain
$\mathbb{H}_t=\mathbb{H}/\gamma_{]0,t]}$, i.e. the upper half plane
from which the curve is removed, to $\mathbb{H}$. In the more general
case of non-simple curves, the function $g_t(z)$ produce conformal
maps from $\mathbb{H}_t=\mathbb{H}/K_t$ to $\mathbb{H}$ where $K_t$ is
the SLE hull at time $t$.  The SLE map $g_t(z)$, where the curve
parametrization $t$ is chosen so that $g_t(z)=z+2t/z+\cdots$ near
$z=\infty$, is a solution of the Loewner equation:
\begin{equation}
\frac{d}{d t} g_t (z)=\frac{2}{g_t(z)-\xi_t} \quad g_{t=0}(z)=z,
\label{Sle_definition}
\end{equation}
where $\xi_t$ is a real valued process, $\xi_t \in \mathbb{R}$, which
drives the evolution of the curve. For a system which satisfies the
Markovian and conformal invariance properties, together with the
left-right symmetry, the process $\xi_t$ is shown \cite{Schramm_2} to
be proportional to a Brownian motion: {\bf E}$[\xi_t]=0$ and {\bf
  E}$[\xi_t \xi_s]= \kappa \,\mbox{min}(s,t)$.  The symbol {\bf
  E}[...]  indicates the stochastic average over the Brownian motion.
The SLE curves are fractal objects and their length, $S$, measured in
units of lattice spacing $a$, scales as a function of the system size
$L$ as $S\sim a(L/a)^{d_f}$ where $d_f$ is the fractal dimension given
by:
\begin{equation}
d_f=1+\frac{\kappa}{8} \; .
\label{fractal_dimension}
\end{equation}
{\it The lattice model and the interface}--- In this letter we
consider the model defined on a square lattice with spin variable
$\sigma_j=\exp{i 2\pi/N n(j)}$ at each site $j$ taking $N$ possible
values, $n(j)=0,1,\cdots,N-1$. The most general $Z_N$ invariant spin
model with nearest-neighbor interactions is defined by the reduced
Hamiltonian \cite{Zama_lat,Dotsi_lat}:
\begin{equation}
H[n]=-\sum_{m=1}^{\lfloor N/2 \rfloor} J_{m}\left[\cos \left(\frac{2\pi m n}{N}\right)-1\right],
\label{reducedH}
\end{equation}
where $\lfloor N/2 \rfloor$ denotes the integer part of $N/2$. The
associated partition function reads:
\begin{equation}
Z=\sum_{\{\sigma\}}\exp\left[-\beta \sum_{<ij>} H[n(i)-n(j)]\right] \; .
\label{partition1}
\end{equation}
For $J_m=J$, for all $m$, one recovers the $N-$state Potts model,
invariant under a permutational $S_N$ symmetry while the case
$J_m=J\delta_{m,1}$ defines the clock model \cite{Potts_Clock}. For
$N=2$ and $N=3$ these models coincide with the Ising and the
three-state Potts model respectively, while the case $N=4$ is
isomorphic to the Ashkin-Teller model \cite{Ashkin,Lin}. Defining the
Boltzmann weights:
\begin{equation}
x_n=\exp\left[-\beta H(n)\right], \quad n=0,1,\cdots,N-1 \; ,
\end{equation}
the most general $Z_N$ spin model is then described by $\lfloor
N/2\rfloor$ independent parameters $x_n$ as $x_0=1$ and
$x_n=x_{N-n}$. The general properties of these models for $N=5,6,7$
have been studied long time ago (see e.g. \cite{Alcaraz2} and
references therein). The associated phase diagrams turn out to be
particularly rich as they contain in general first-order, second-order
and infinite-order phase transitions.  For all the $Z_N$ spin models
it is possible to construct a duality transformation (Kramers-Wannier
duality). In the self-dual subspace of
(\ref{reducedH})-(\ref{partition1}), which also contains the Potts and
the clock model, the $Z_N$ spin model are critical and completely
integrable at the points \cite{Zama_lat2, Alcaraz}~:
\begin{eqnarray}
x^{*}_0= 1 \; ; \; x^{*}_n&=&  \prod_{k=0}^{n-1} \frac{\sin \left(\frac{\pi k}{N}+\frac{\pi}{4
    N}\right)}{\sin \left(\frac{\pi (k+1)}{N}-\frac{\pi}{4 N}\right)} \; .
\label{integrablecond}
\end{eqnarray}
There is strong evidence that the self-dual critical points
(\ref{integrablecond}), referred usually as Fateev-Zamolodchikov (FZ)
points, are described in the continuum limit by $Z(N)$
parafermionic theories \cite{Alcaraz3}.  Very recently, a further
strong support for this picture has been given in \cite{Rajabpour}
where the lattice candidates for the chiral currents generating the
$Z_N$ symmetry of the continuum model has been constructed.

Consider now the model, at the self-dual critical point, defined on a
simple connected domain.  By choosing some specific boundary
conditions, for each spin configuration there is a domain wall
connecting two fixed points on the boundaries (see below for some
specific example).  In general one is interested in the conformally invariant
boundary conditions which, for a given bulk CFT, represent a finite
set into which, under renormalization group, any uniform boundary
condition will flow \cite{Cardy_review_bcft}.  The change of
conformally boundary conditions at some point of the boundary is
implemented in CFT by the insertion at that point of a given boundary
conditions changing (b.c.c.) operator \cite{Cardy_review_bcft}.

By carefully choosing the boundary conditions, the associated domain
wall connecting the two points at the boundaries is then expected to
be described by measures which are invariant under conformal
transformation. This can be understood from the fact that the
expectation values describing the curve correspond in the continuum
limit to the correlation functions of the CFT with the insertion of
the two b.c.c operators.

In order to establish the SLE/CFT connection, the b.c.c. operator
associated to the interface have to satisfy particular relations under
the action of the symmetry generators, the so-called null state
condition.  In \cite{Raoul} the existence of such operators in the
$Z(N)$ parafermionic was pointed out. One of these b.c.c. operator,
inserted at a point $x_0$, generate the condition where the spins are
fixed to (say) the value $A$ on the left side of $x_0$ while they can
take the other $N-1$ values $B,C,..$ with equal probability on the
right side (in the following we indicate the possible values of the
spins with the letters $A, B\cdots$).  Interpreting the b.c.c. null
state condition via the introduction of an additional stochastic
motion in the $Z_N$ internal space independent from
(\ref{Sle_definition}), the geometric property of the interface
generated by such boundary conditions was predicted to be described
for $N\geq 4$ by an SLE with $\kappa=4(N+1)/(N+2)$ \cite{Raoul}, thus
the prediction
\begin{eqnarray}
d_f= 1 + \frac{1}{2} \frac{(N+1)}{(N+2)} \; .
\label{df}
\end{eqnarray}
We will test this relation in the following.

{\it Numerical simulation}
\begin{figure}[ht]
\begin{center}
\epsfxsize=260pt\epsfysize=260pt{\epsffile{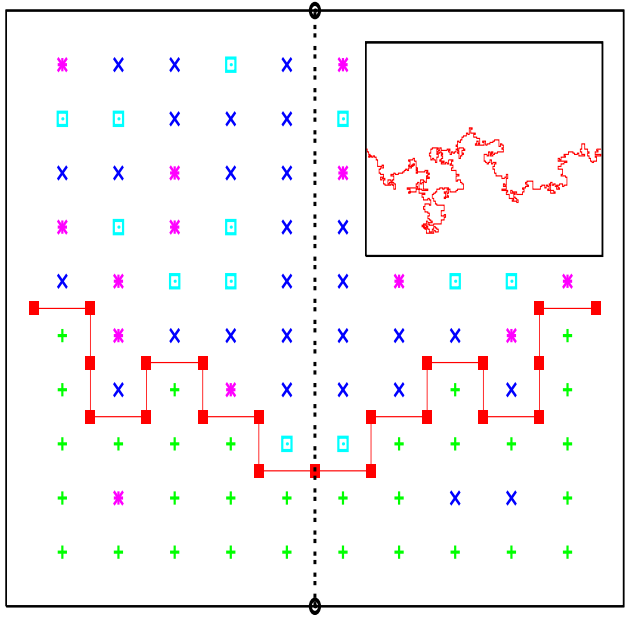}}
\end{center}
\caption{
Definition of the interface. The interface separetes the spins with a fixed value connected to the
bottom boundary from spins with other colors.
The vertical dashed line corresponds to 
the test of the crossing probabilities against Schramm's formula.
The inset contains a typical 
$320 \times 320$ configuration for the $Z(5)$ parafermionic theory.
\label{interface}
}
\end{figure}
Our goal is to compute the interface and check the validity of
eq.(\ref{df}) for the two cases $N=4$ and $N=5$ which are the
simplest lattice models described by non-minimal conformal field
theories.  

We are going to compute the fractal dimension associated to the
interface which crosses the lattice. To create this interface, we
impose that half of the spins on the boundary take a fixed value $A$,
these spins being connected two by two, while the remaining boundary
spins are forced to take values different from $A$. Then the interface
will be the border of the geometric cluster of spins taking a value
$A$ and connected to the spins on the boundary with fixed spins. We
show an example of such a configuration in Fig.~\ref{interface}. 
In this figure, the spins with a fixed value are the ones which touch
the bottom boundary. The interface is shown as the line which connects
the left boundary to the right boundary.
Similar conditions were considered in a
recent work by Gamsa and Cardy for the $Q=2$ and $Q=3$ Potts model
case\cite{Gamsa} who obtained a good agreement with the prediction of
the corresponding formula (\ref{fractal_dimension}) for the Potts
models. This type of boundary condition, which was called fluctuating
in \cite{Gamsa}, ensures that there is a unique interface which
crosses the lattice. We should also mention that on the square
lattice, the definition of the interface can contain some
ambiguities. There are different ways of dealing with these
ambiguities but the large size results will not be affected by
them\cite{MPRS}.  For the simulation of the $Z(4)$ and $Z(5)$ model at
the FZ point, we employed a standard Monte Carlo algorithm. One can
also use a cluster algorithm but with the boundary conditions that we
consider, it turns out to be less efficient than Monte Carlo. We
performed simulations on square lattices of rectangle geometry $L_x
\times L_y$, the interface being created along the $y$ direction. We
simulate the size $L_x = 10,20,30,40,60,80$ and $160$ with for each
size $L_y = L_x$ and $L_y = 3 L_x$. For the larger linear size $L_x$
that we consider, we see very little difference between these two
cases.
\begin{figure}[ht]
\begin{center}
\epsfxsize=260pt\epsfysize=260pt{\epsffile{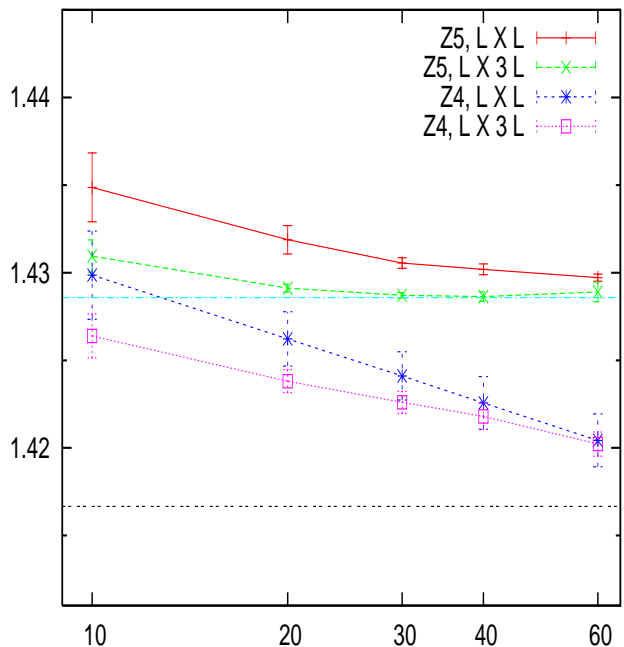}}
\end{center}
\caption{
$d_f$ vs. $L_{min}$ for $N=4$ and $N=5$. The straight lines correspond to the predictions 
of eq.(\ref{df}).
\label{Z4Z5}
}
\end{figure}

For $L_x=L_y$, we simulated 1 million independent configurations up to
$L_x=80$ and $200 000$ configurations for $L_x=160$. For $L_y = 3
L_x$, we simulated 1 million in-dependant configurations up to
$L_x=60$, $300 000$ configurations for $L_x=80$ and $50 000$
configurations for $L_x=160$. The autocorrelation time grows as $\tau
\simeq L^z$ with $z \simeq 2.2(1)$ for both $N=4$ and $N=5$ and for
the two ratios $L_y/L_x$ that we considered. For the largest sizes, we
obtain $\tau \simeq 8000$ for $L_y = L_x$ while $\tau \simeq 15 000$
for $L_y = 3 L_y$ for both $N=4$ and $N=5$.

Fig.~\ref{Z4Z5} contains our main results. In this figure, one shows
the exponent $d_f$ obtained by doing a fit of $S \simeq L_x ^{d_f}$
with data in the range $L_x=[L_{min}, \cdots , 160]$. For $N=4$, the
measured value moves close the predicted value $d_f= 1 + 5/12$. The
deviation for the larger size that we can measure is of order 1/4 \%
and from the figure, we expect that this deviation will decrease for
larger size.  For $N=5$, the agreement is already perfect for the
larger sizes and for $L_y = 3 L_x$. For $L_y = L_x$, there is still a
small deviation (of order 1/10 \%) but again this deviation decrease
while increasing the size. The fact that the agreement is better for
$N=5$ than for $N=4$ is not surprising since the $Z(4)$ parafermionic
field theory has a $c=1$ central charge. CFTs with such a central
charge are known to contain marginal operators which may produce
strong finite size effects.

Further tests can also be done like in \cite{Gamsa,BLDM}. These
authors made additional checks like the test against Schramm's formula
or the computations of $\kappa$ from the statistics of the Loewner
driving function obtained by ``unfolding'' the interfaces. Actually,
for our purposes, these measurements turn out to be not very practical
and precise due to the fluctuating boundary conditions and the
geometry that we employed. Indeed, concerning Schramm's formula, these
boundary conditions explicitly breaks the $Z_N$ symmetry and the
left-right symmetry is expected to be recovered only in the very large
scale limit. One observes then strong finite size corrections as
already observed by Gamsa and Cardy for the $Q=3$ Potts model with the
same type of boundary conditions. Note that in our case we have more
states (4 or 5) and thus the boundary conditions are even more
asymmetrical. We tested crossing probabilities against the Schramm's
formula along the line indicated in Fig. 1. The best fit gives a value
of $\kappa=3.41(2)$ for $Z(4)$ and $\kappa=3.42(2)$ for $Z(5)$ which
is close to the expected results. The agreement in both cases of the
numerical data compared to Schramm's formula is of order $1\%$ which is
comparable to the result in \cite{Gamsa}. Concerning the direct
extraction of $\kappa$ the situation is even worse since the unfolding
transformation is singular. To bypass the problem, one should use a
different geometry. For the $Q=3$ Potts model, the disk geometry on
the triangular lattice was suitable and provided good results
\cite{Gamsa}.  This configuration is not possible in our case since
the location of the critical point is not known for the triangular
lattice. In this respect we mention that a method to find these
critical points for $Z(N)$ spin models on different lattices has been
proposed in \cite{Rajabpour}.

In this letter we obtained the first results on the geometry on the
interfaces which are expected to be described by SLE in non minimal
CFTs. We provide strong numerical support to the validity of the
exponent eq.(\ref{df}) first obtained in \cite{Raoul}. The agreement
is excellent for both cases that we considered with $N=4$ and $N=5$.
We believe that these results give support to the theoretical approach
proposed in \cite{Ludwig2,Raoul} to describe non minimal CFTs by
SLE with additional stochastic motion in the internal degrees of freedom.

\end{document}